\documentclass[pre,twocolumn,aps,floatfix,superscriptaddress]{revtex4}

\usepackage[english]{babel}
\usepackage{amssymb,amsfonts,amsmath}
\usepackage[pdftex]{graphicx}
\usepackage{latexsym}
\usepackage{multirow}
\usepackage{color}

\usepackage{amsmath}
\usepackage{hyperref}

\begin{document}

\title{Link aggregation process for modelling weighted mutualistic networks}

\author{Manuel Jim\'enez-Mart\'{\i}n}
\affiliation{Departamento de F\'{\i}sica Fundamental, Despacho 2.27, Facultad de Ciencias, Universidad Nacional de Educaci\'{o}n a Distancia (UNED), Paseo Senda del Rey 9, E-28040, Madrid,  Spain. +34 91 3987636.}

\author{Juan Carlos Losada}
\author{Juan Manuel Pastor}
\author{Javier Galeano}
\affiliation{Grupo de Sistemas Complejos, Dept. Ciencia y Tecnolog\'{\i}a aplicadas a la I.T. Agr\'{\i}cola, E.U.I.T. Agr\'{\i}cola. Universidad Polit\'ecnica de Madrid (UPM). Ciudad Universitaria, s/n, 28040, Madrid, Spain }

\begin{abstract}

Mutualism is a biological interaction mutually beneficial for both species involved, such as the interaction between plants and their pollinators. Real mutualistic communities can be understood as weighted bipartite networks and they present a nested structure and truncated power law degree and strength distributions. We present a novel link aggregation model that works on a strength-preferential attachment rule based on the Individual Neutrality hypothesis. The model generates mutualistic networks with emergent nestedness and truncated distributions. We provide some analytical results and compare the simulated and empirical network topology. Upon further improving the shape of the distributions, we have also studied the role of forbidden interactions on the model and found that the inclusion of \emph{forbidden links} does not prevent for the appearance of super-generalist species. A Python script with the model algorithms is available.

\end{abstract}

\maketitle







\section{Software Availability}
The algorithm for the Mutualistic Link Model and several analysis functions are available on a Python script downloadable from our webpage.\\
Name of software: Mutualink.\\
Developer: Manuel Jim\'enez-Mart\'{\i}n.\\
Language: Python 2.7 (requires numpy, scipy, pylab and matplotlib modules).\\
Contact address: Dept. Ciencia y Tecnolog\'{\i}a aplicadas a la I.T. Agr\'{\i}cola, E.U.I.T. Agr\'{\i}cola. Universidad Polit\'ecnica de Madrid. Ciudad Universitaria, s/n, 28040, Madrid. Spain\\
E-mail address: manuel.jimenez@bec.uned.es\\
URL: \url{ http://hypatia.agricolas.upm.es/JJ_lab/mutualink.html}\\

\section{Introduction}

It is estimated that close to 40\% of angiosperm plants, from crops to tropical trees, are self-incompatible and depend on mutualistic interactions with animals to complete their life cycles \cite{Kohn2008,Teja2006}. In these ecological interactions such as the interplay between plants and their pollinators or seed-dispersers, both involved species obtain mutual benefits \cite{Jordano2000, Bascompte2009}. The essential role of these mutualistic interactions in the biodiversification and coevolution processes has been discussed in depth in the ecological literature \cite{Bastolla2009}. Although exclusive pairwise interactions between two species exist in nature, the majority of species distribute their interactions among several different mutualistic partners. Thus, most ecosystems exhibit an intricate pattern of mutualistic relationships among their species that can be represented as a bipartite complex network with two disjoint sets of nodes representing animal and plant species, $A$ and $P$. The complex network approach allowed the study of mutualism at community level, revealing the enormous relevance of community structure onto the ecological and coevolutionary processes of ecosystems. The structure and properties of mutualistic networks, such as nestedness and truncated distributions, have been thorougly studied during the last decade \cite{Bascompte2003, Jordano2003}. 

Mutualistic communities can be studied as weighted bipartite networks where interactions between species are represented as weighted links between $A$ and $P$ nodes. All network information is included in the quantitative interaction matrix $\textbf{W}$ where element $w_{ij}$ represents the observed frequency of interaction between species $A_i$ and $P_j$ \cite{Gilarranz2012,Olesen2011}. Species can be characterized by their degree $k$ -i.e. the number of species with whom they interact- and by their strength $s$ -i.e. the sum of the weights of their links: $s_i^A=\sum_j w_{ij}$ and $s_j^P=\sum_i w_{ij}$. Both magnitudes are correlated in mutualistic networks \cite{Vazquez2003,Gilarranz2012} meaning that species with a high number of interaction partners interact more frequently than less connected species. Species with high $k$ and $s$ values are called \emph{generalists} as opposite to those with lower $k$ and $s$ which are called \emph{specialists}.

Mutualistic networks exhibit scale-free behaviour with cut-offs in both degree and strength distributions for both classes of nodes \cite{Jordano2003,Gilarranz2012}. This means that there is not one big \emph{hub} or \emph{super-generalist} that would accumulate the majority of interactions of the network. This role is instead diluted among several generalist species. Given the fact that mutualistic networks have a few hundreds of nodes at most,  truncations have been considered to be finite size effects on an otherwise scale-free topology \cite{Amaral2000}. An alternative interpretation for the cause of the truncations is the existence of \emph{forbidden links} \cite{Jordano2003}: It is well known that morphological or phenological constraints among species limit the number of interactions observed in the network \cite{Olesen2011}. Regarding the connectivity pattern, generalists tend to interact strongly with generalists forming a dense \emph{core} of interactions that acts as the backbone of the network. Most specialists interact mainly with generalists from the \emph{core} while specialist-specialist interactions are rare and weak. The generalist \emph{core}  provides alternative paths between species and allows the network to remain functional upon extintions of specialists. Remarkably, interaction partners of any given $A$ or $P$ species usually fall into a subgroup of the interaction partners of the next more connected species. This pervasive architecture is called \emph{nestedness} \cite{Vazquez2003,Bascompte2006}. Ordering the rows and columns of $\textbf{W}$ from generalist to specialist, nestedness appears as a triangular-like pattern of interactions that can be expressed mathematically with the condition

\begin{equation}\label{eq:nest}
w_{ij} \geq \textnormal{max} (w_{i+1,j},w_{i,j+1}). 
\end{equation}

There are several metrics to quantify nestedness and the nested pattern have been found to be statistically significant for the vast majority of recorded cases \cite{AlmeidaNeto2008,Galeano2009,Almeida-Neto2011,Allesina2012}. Truncated distributions and nestedness constitute the two main features of mutualistic networks and have been reported in mutualistic communities for every kind of ecosystem, from arctic tundra to tropical forest. While the benefits of mutualistic network architecture are well understood, the mechanisms leading to the formation of mutualistic networks are still the subject of speculation \cite{Medan2007,Guimaraes2007,Maeng2012}. 


Considering information on species population gives new insight. It is known that species degree, strength and abundance are correlated in mutualistic communities \cite{Vazquez2003, Gilarranz2012}. A high degree implies higher reproductive benefits for the species which translates into a higher species population and therefore more frequent interactions, leading to a higher species strength. Taking account of this fact, the \emph{Individual Neutrality hypothesis} \cite{Vazquez2005,Vazquez2005a,Vazquez2005b} affirms that mutualistic network topology can be explained to a great extent by considering population information and random interaction between individuals. This means that interaction probability is considered to be proportional to species abundance, ignoring any kind of morphological or phenological preference between interacting species. It was recently shown that quantitative nestedness can be explained as a mere mass effect \cite{Allesina2012} considering species population. It is the aim of this article to extend this idea showing that both nestedness and truncated distributions may emerge from simple rules considering population effects.

In this article we present a network growth model based on individual neutrality that succesfully generates realistic mutualistic networks exhibiting both truncated distributions and nestedness. Shifting the focus from species towards individuals ultimately means to describe the network growth at the level of individual links. In contrast to previous node aggregation binary models, our model constructs weighted networks by means of a unitary link aggregation process. The article is structured as follows: Firstly, the model description along with some analytical results are presented on section \ref{sec:model}. We have then compared the simulated network topology with 9 empirical datasets on section \ref{sec:results}. In section \ref{sec:FL} we study the effect of forbidden links on the model. Finally, section \ref{sec:conc} is devoted to the conclusions.

\begin{center}
\begin{figure*}[t]
\includegraphics[width=\linewidth]{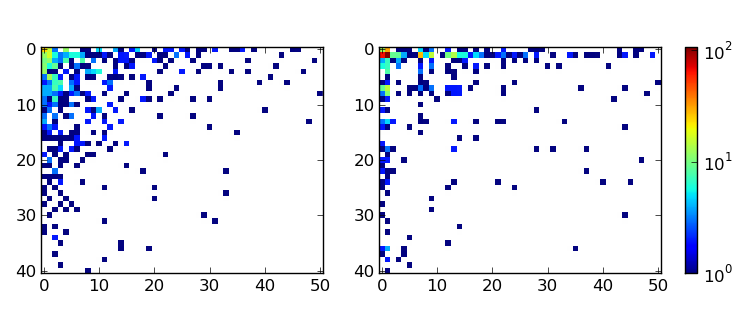}  
\caption{{\bf Empirical (left) and simulated (right) weighted interaction matrices} for the network from \cite{Bluthgen2004} with parameters $W=641$, $N_A=41$, $N_P=51$. Rows and columns represent animal and plant species respectively. Element color represents link weight i.e. frequency of the interaction. Absent links are displayed in white.}
\label{fig:sample}
\end{figure*}
\end{center}


\section{Materials and Methods} 
\label{sec:model}

On the basis of the individual neutrality hypothesis, our model simulates the formation of mutualistic networks as a dynamical link aggregation process. Our main assumption is that every individual on the ecosystem participates on just one single interaction with an individual of the opposite class. Such unitary interactions are incorporated to the network as a single link of weight equal to $1$. Consequently, on our model species population is equal to species strength $s$. Thus, the individual neutrality hypothesis resut on a strength-preferential attachment process, since the interaction probability is proportional to species population, i.e. strength. Both end-nodes of an incoming link are selected independently with probabilty proportional to each species strength. It is also possible to select a disconnected node (species with strength zero) with probability given by some function $p_{\alpha}^{\textnormal{new}}$. This accounts for the arrival of a new species of class $\alpha$ ($A$ or $P$) into the ecosystem. Every time-step a unitary link is added to the network so the number of links acts as a time-variable: $w(t)=\sum_{ij} w_{ij}(t)=t$.

Only three input parameters are needed: number of animal species $N_A$, number of plant species $N_P$ and total network weight $W$. The algorithm is as follows.  Initially, all of the $N_{A}·N_{P}$ possible interactions have weight zero. Starting with a completely connected seed of 4 nodes (2 of each class), every time-step $t$ an unitary interaction is incorporated to the network according to the following rules:

\begin{enumerate}
\item Firstly we check for the appearance of a new species of each class in the ecosystem. A disconnected node will be selected with probability $p_{\alpha}^{\textnormal{new}}(t)$, where $\alpha$ stands for classes \emph{A} and \emph{P}. 

\item Otherwise, an already connected $\alpha$ node will be selected via strength-preferential attachment with probability $(1-p_{\alpha}^{\textnormal{new}})p_{\alpha}^{i}(t)=s_{i}(t)/w(t)$.

\item After determining both end-nodes of the incoming link, the weight of the resulting $(A_{i},P_{j})$ interaction is increased in one unit.
\end{enumerate}

The process is repeated until the final weight of the network is reached, $t=W$. The final size of the network $N_{A}N_{P}$ and its topology depends on the specific functional form of the \emph{birth} functions $p_{\alpha}^{\textnormal{new}}$. It seems reasonable to consider the appearance of a new species being less likely as the number of species grows due to limited resources and space. For the sake of simplicity we have considered the following form:

\begin{equation}\label{eq:birth_prob}
p_{\alpha}^{\textnormal{new}}(t)=\frac{\lambda_{\alpha}}{N_{\alpha}(t)} \: ,
\end{equation}

\noindent where the $\lambda_{\alpha}$ are parameters. Note that $p_{\alpha}^{\textnormal{new}}$ only depends on the current number of species of its own class, so the growth processes for each class are uncoupled. This is of course a heavy assumption. Certainly, there are many conceivable choices for the \emph{birth} functions. This specific form is simple enough to draw analytic results and it produces network topologies similar to those observed in real mutualistic communities.\\

The $\lambda_{\alpha}$ weights can be expressed as a function of the network parameters and allow to simulate networks of any desired size $N_{A}·N_{P}$. The necessary time for $N_{\alpha}(t)$ to increase one unit is $\Delta t=N_{\alpha}(t)/\lambda_{\alpha}$. Summing up to number of species $N_{\alpha}(t)$ and weight $w(t)=t$ we arrive to the following expression

\begin{equation}\label{eq:lambda}
\lambda_{A}=\frac{N_{\alpha}(t)(N_{\alpha}(t)-1)}{2t}.
\end{equation}

This relationship holds at every time-step. Substituting the network parameters we finally get:
 
\begin{equation}
\lambda_{A}=\frac{N_{A}(N_{A}-1)}{2W}\: ; \:\lambda_{P}=\frac{N_{P}(N_{P}-1)}{2W} \; .
\end{equation}

\begin{center}
\begin{table*}[!bth]
\resizebox{\linewidth}{!}{
\begin{tabular}{|rlcccccccc|}
     \hline
     & Network & $W$ & $E$ & $N_{A}$ & $N_{P}$ & $\lambda_{A}$ & $\lambda_{P}$ & $M$ & $C$ \\
     \hline
     N1 \cite{Bluthgen2004}&Bl\"{u}thgen et al. (2004) &644&285&41&51&1.27&1.98&2091&0.14\\ \hline
     N2 \cite{Memmot1999}&Memmott (1999) &2183&299&79&25&1.41&0.14&1975&0.15\\ \hline
     N3 \cite{Barrett1987}&Barret \& Helenurm (1987) &550&167&102&12&9.37&0.12&1224&0.14\\ \hline
     N4 \cite{Elberling1987}&Elberling \& Olesen (1999) &383&238&118&23&18.02&0.66&2714&0.09\\ \hline
     N5 \cite{Kato1990}&Kato et al. (1990) &2384&1202&678&89&96.27&1.64&60342&0.02\\ \hline
     N6 \cite{Inouye1988}&Inouye \& Pyke (1988) &1459&281&91&42&2.81&0.59&3822&0.07\\ \hline
     N7 \cite{Schleuning2010}&Schleuning et al. (2010) (1) &3447&419&88&33&1.11&0.15&2904&0.14\\ \hline
     N8 \cite{Schleuning2010}&Schleuning et al. (2010) (2) &3081&288&71&15&0.81&0.03&1065&0.27\\ \hline
     N9 \cite{Schleuning2010}&Schleuning et al. (2010) (5) &2802&283&71&19&0.89&0.06&1349&0.21\\ \hline
\end{tabular}}
\caption{
{\bf Empirical networks parameters.} Network weight $W=\sum_{ij}w_{ij}$, number of links $E=\sum_{ij}A_{ij}$ (where $A_{ij}$ is the standard adjacency matrix with entries either 0 or 1), 		number 	of animal and plant species $N_{\alpha}$, birth probability weight $\lambda_{\alpha}$, magnitude $M=N_{A}·N_{P}$ and connectivity $C=E/M$.}
\label{tab:tab1}
\end{table*}
\end{center}

\begin{figure}[!h]
	\begin{center}
	\includegraphics[width=\linewidth]{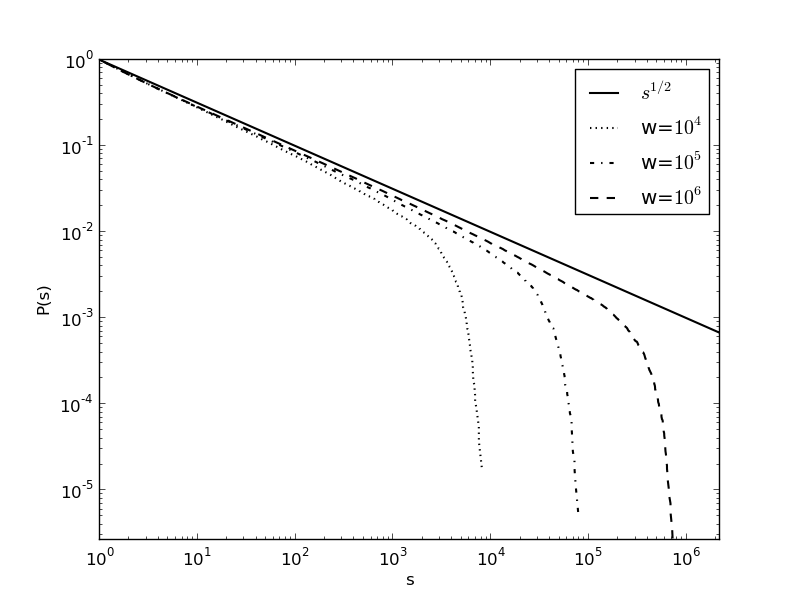}   
	\end{center}
  	\caption{
  	{\bf Large networks cumulative strength distribution} for both classes of nodes for a symmetric network of magnitude $M=2$ and increasing $w$. The theoretical limit of $-0.5$ for infinite networks is shown as well. 
  	}
  	\label{fig:thExp}
\end{figure}

Since the node birth process is equivalent for both classes, in the subsequent we will drop the $\alpha$ subindices. Also, we will express the number of species at time t as $N_{t}$. Now, we will prove that our model generates strength scale-free strength distributions for large enough networks. The change in the number of nodes of strength $s$ at time $t$ is equal the number of nodes with strength $s-1$ receiving a link minus the number of nodes of strength $s$ receiving a link. The rate equation is

\begin{equation}\label{rate}
N(t+1)P(s,t+1)-N(t)P(s,t)=
\frac{s-1}{t}N(t)P(s-1,t)-\frac{s}{t}N(t)P(s,t) \: ,
\end{equation}

where $N(t)$ is the number of nodes the corresponding class present in the network at time $t$ and $P(s,t)$ is the node probability for having strength \emph{s} at time \emph{t}. Using equation \ref{eq:lambda}, we can express the number of as

\begin{equation}\label{eq:N(t)}
N(t)=\frac{1+\sqrt{1+8\lambda t}}{2}\simeq\sqrt{2\lambda t} \: 
\end{equation} 

when $t\rightarrow \infty$. Assuming that the strength probability distribution will reach stationarity for large enough times we can approximate $P(s,t+1)\cong P(s,t)\equiv P(s)$ and arrive to the following recurring relation

\begin{equation}\label{eq:N(t)2}
P(s)=\frac{s-1}{s+\xi(t)}P(s-1) \:,
\end{equation}  

where $\xi(t)=\sqrt{t^2+t}-t=\frac{1}{2}+O(\frac{1}{t})$. Developing equation \ref{eq:N(t)2} we finally arrive to a power law expression for the strength distribution.

\begin{equation}\label{eq:N(t)3}
P(s)=\frac{\Gamma(s)}{\Gamma(s+3/2)}P_{1,t}\sim s^{-3/2} \:,
\end{equation}  

where we have used the following relationship for Euler Gamma functions: $\Gamma(z+a)/\Gamma(z+b)=z^{a-b}+O(z^{a-b-1})$ \cite{Tricomi1951}. Hence, the individual interaction aggregation process with strength-preferential attachment generates networks with scale-free strength distribution in the limit of large networks. A known result for power-law distributions  states that the exponent of the cumulative distribution must be $1$ unit larger \cite{Clauset2009}, giving an exponent of $-0.5$ for the cumulative strength distribution. Figure \ref{fig:thExp} shows the shape of the cumulative strength distribution of increasingly large networks tending towards the theoretical value of $-0.5$ and the truncations observed on the finite network strength distributions are finite size effects.

\section{Results and Discussion} 

\subsection{Simulation results} \label{sec:results}
We have studied the topology of 9 empirical networks as well as their simulated counterparts generated by the model. The datasets depicted in table \ref{tab:tab1} were extracted from the Interaction Web Database \ref{IWDB}. For each real network we run the model $500$ times with the corresponding input parameters $W$, $N_A$ and $N_P$. We then computed the average nestedness, as well as the average strength and degree cumulative distributions.\\

\begin{table}[!bt]
\resizebox{\linewidth}{!}{
\begin{tabular}{|r|ccccccccc|}
	\hline
	NODF & N1 & N2 & N3 & N4 & N5 & N6 & N7 & N8 & N9 \\ \hline
	Empirical & 43.51 & 42.84 & 30.78 & 15.28 & 7.67 & 17.22 & 34.58 & 51.15 & 45.83 \\
	Simulated & 49.06 & 65.80 & 57.99 & 40.22 & 28.69 & 55.76 & 65.66 & 67.47 & 67.64 \\ \hline
\end{tabular}}
\caption{{\bf NODF nestedness index for the empirical and simulated networks.} Results for the simulated networks where calculated as the average index after 500 runs of the model. Both empirical and simulated networks exhibit highly significative nestedness with $p<0.01$. The p-values were calculated after 500  randomizations of the interaction matrix. Each randomization was performed by randomly shuffling the non-zero matrix elements.}
\label{tab:TNestW}
\end{table}

The standard \emph{NODF} metric \cite{AlmeidaNeto2008} was computed to quantify nestedness (table \ref{tab:TNestW}). Along with it, \emph{p}-values were calculated in order to assess the statistical significance of the results against random permutation of matrix elements. All networks, both real and simulated, exhibit highly significative nestedness, $p<0.01$, although index values are always higher for simulated networks. The highly nested structure of the simulated networks is a consequence of the strength-preferential attachment rule. In our model, nestedness emerges from simple interaction rules: random interactions between individuals and interaction probability proportional to species population. Despite the qualitative success there are differences between real networks and the networks generated by the model. Compared to empirical mutualistic networks, simulated networks are closer to perfect nestedness, i.e. a higher number of links fulfill the nestedness condition expresed by equation \ref{eq:nest}. The simulated networks exhibit a very compact generalist \emph{core} where basically every possible interaction between generalist species is present. On the other hand, real mutualistic networks show a more dispersed \emph{core} with some holes and slightly smaller link weights. Real networks are also noisier and they show a higher number of specialist-specialist interactions that are not present on the simulated networks. This is better understood examining the degree and strength distributions. \\ 
  
\begin{figure}[!h]
	\begin{center}
	\includegraphics[width=\linewidth]{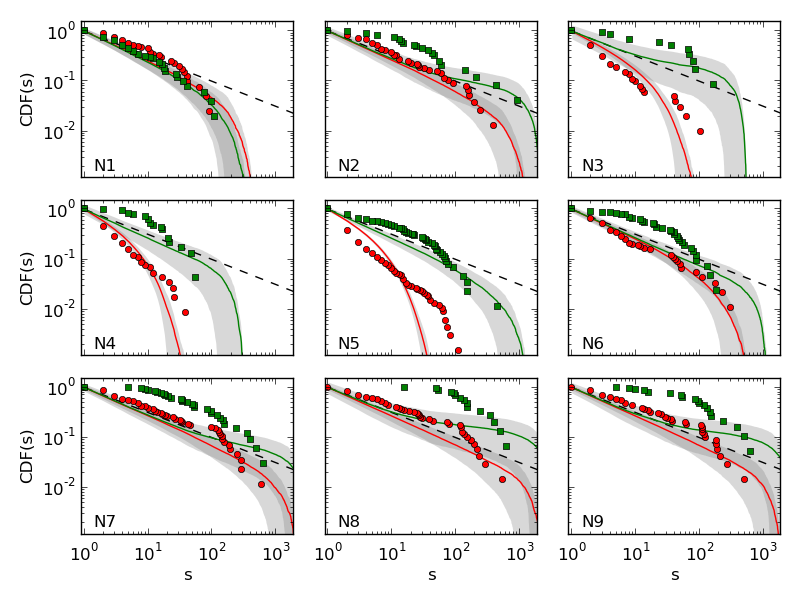}   
	\end{center}
  	\caption{
  	{\bf Cumulative strength distributions for the 9 networks studied.} Red circles and green squares correspond to empirical $A$ and $P$ distributions respectively. Solid red and green lines are the average $A$ and $P$ distributions after 500 runs of the model with the corresponding parameters. Gray regions represent the standard deviation from the average. Black dashed lines correspond to the exponent of $0.5$ predicted for infinitely large networks. Model distributions of networks 7 ($W=3447$),8 ($W=3081$) and 9 ($W=2802$) fit well the theoretical exponent in their power law regions.}
  	\label{fig:Sdist}
\end{figure}

\begin{figure}[!h]
	\begin{center}
	\includegraphics[width=\linewidth]{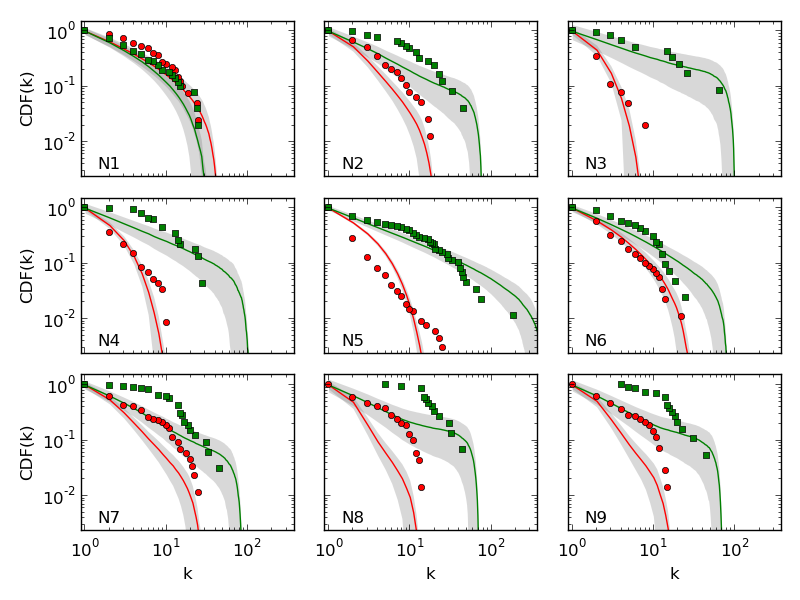}   
	\end{center}
  	\caption{
  	{\bf Cumulative degree distributions for the 9 networks studied.} Red circles and green squares correspond to empirical $A$ and $P$ distributions respectively. Solid red and green lines are the average $A$ and $P$ distributions after 500 runs of the model with the corresponding parameters. Gray regions represent the standard deviation from the average.  Model distributions show emergent exponential truncations although node degree is not a variable used by the model.}
  	\label{fig:Kdist}
\end{figure}

Figures \ref{fig:Sdist} and \ref{fig:Kdist} represent the average cumulative strength and degree distributions for the simulated networks along with the real data. The distributions are shown separately for each class $A$ and $P$. Exponential truncations are observed for the simulated strength distribution. These are merely a consequence of the finite size of the system since the model generates scale-free strength distributions in the limit of large networks as demonstrated in the previous section. It can be observed how the simulated distributions approach the theoretical exponent on $0.5$ on the heavier networks (networks $7$, $8$ and $9$). Remarkably, although node degree does not come into play in the network growth process, an exponential decay also appears in the degree distributions. This supports the hypothesis that the well-known binary network topology may arise from random individual encounters. However, there exist some discrepancy with the empirical datasets. For instance, cut-offs are generally higher on the simulated networks for both strength and degree distributions while the middle section of the distributions usually rest at lower values than the empirical ones, specially for the $P$ nodes (plants). Very low probability points at high strength and degree values correspond to a super-generalist species receiving a large number of connections with very high weights. This super-generalists are effectively stealing interactions from less connected species, leading to less and weaker interactions among specialists than those observed on the empirical networks. In the absence of any limiting mechanism, super-generalist arise as a consequence of the strength-preferential attachment process. \\

Despite the model success at qualitatively reproducing nestedness and truncated $s$ and  $k$ distributions, the results are far from a quantitative match with the data from real networks. Two main disagreements between simulated and empirical networks have been pointed out: \emph{a)} simulated networks exhibit a more nested structure than empirical oness, with fewer interactions not fulfilling the nested condition \ref{eq:nest}, and \emph{b)} simulated networks have broader $s$ and $k$ distributions than empirical ones, with higher values of $s_{max}$ and $k_{max}$ and less number of interactions in the medium range. Both this facts are a consequence of the strength preferential attachment. Since there is no restriction on the number of links a species can receive, a super-generalist node of each class appears, taking a great amount of interactions in detriment of other less connected species, thus pushing the cut-off of the distributions to higher values. This also results in a smaller but stronger \emph{core} than observed in real networks (figure \ref{fig:sample}). 
  	
\subsection{Forbidden Links} \label{sec:FL}
Some authors have suggested forbidden interactions as a plausible explanation for nestedness and truncated distributions. We have studied the effect of forbidden links on the topology of the model generated networks. The objective was to test the hypothetical role of forbidden links as a limiting mechanism that could prevent the appearance of supergeneralist species,  lowering the distribution's cut-offs and improving nestedness values.\\
\begin{figure}[!h]
	\begin{center}
	\includegraphics[width=\linewidth]{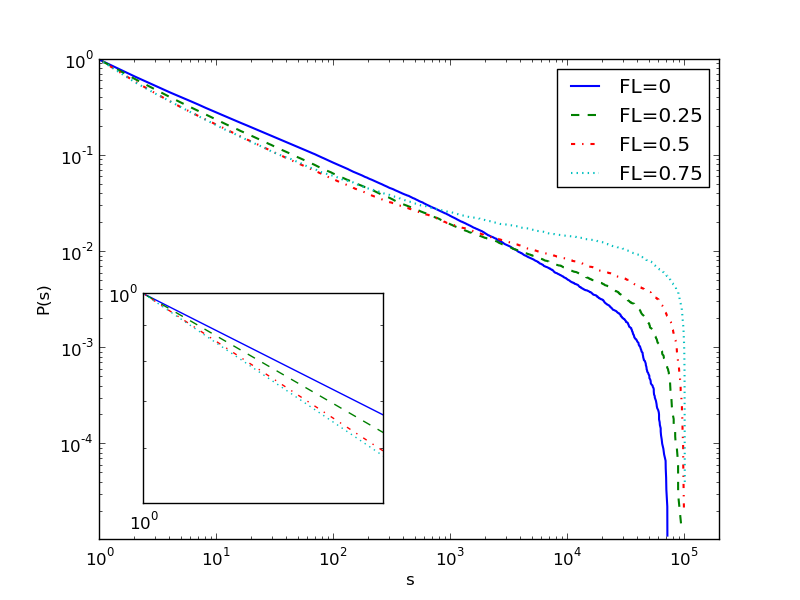}   
	\end{center}
  	\caption{
  	{\bf Joint cumulative strength distribution for different $FL$ percentages.} Distributions were obtained as the average of 500 simulations with parameters $W=10^5$, $M=2$. An increase of  $s_{max}$ is observed. The inset shows the decrease of the power law exponent at low $s$ values. These facts imply a loss of strength for mildy connected species in favour of the supergeneralists.
  	}
  	\label{fig:FLS}
\end{figure}

The implementation of forbidden links on the model is straightforward. A percentage $FL=\frac{m}{N_AN_P}$ of fixed forbidden interactions is selected a priori by randomly choosing $m$ forbiden elements on matrix $\textbf{W}$. Then, on algorithm step 3, once both end-nodes of the incoming interaction have been selected we test if the resulting interaction is forbidden. If this is the case, the current iteration is discarded and the process is repeated from step 1. Note that if the iteration is discarded the total weight of the network remains the same and the time variable is not increased.\\ 

\begin{figure}[!h]
	\begin{center}
	\includegraphics[width=\linewidth]{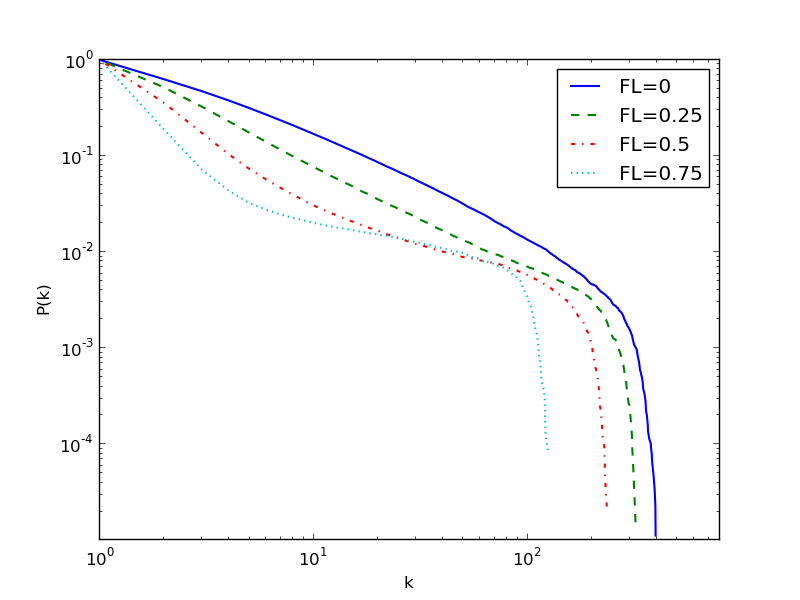}   
	\end{center}
  	\caption{
  	{\bf Joint cumulative degree distribution for different $FL$ percentages.} Distributions were obtained as the average of 500 simulations with parameters $W=10^5$, $M=2$. The decrease in $k_{max}$ implies a loss of links that ultimately damages the network topology.   
  	}
  	\label{fig:FLK}
\end{figure}

The percentage of unobserved links on empirical mutualistic communities datasets can be as high as 80\% of the $N_AN_P$ potential interactions \cite{Olesen2011}. To explore the effect of forbidden links we run simulations on a large network to avoid finite size effects to a certain extent. We would expect an exponential cut-off on the scale free distributions caused by the forbidden links. Figure \ref{fig:FLS} shows the cumulative strength distributions for increasing values of $FL$. On the contrary to expected, introducing forbidden links not only does not lower the value of $s_{max}$ but it slightly increases it. Consequently forbidden links do not prevent from the appearance of super-generalists. However, it does increase the power law exponent for low values of $s$, pushing downwards the middle section of the distribution. As we move towards high $FL$ percentages the effect becomes more evident and the networks topology changes significatively. The shape of the resulting distribution correspond to networks with a high number of nodes receiving few low-weight links and very few strongly connected species receiving most of the network connections. The corresponding degree distributions are represented on figure \ref{fig:FLK}. The power law exponent decreases noticeably with $FL$ and the value of $k_{max}$ lowers drastically with $FL$. This is the consequence of a net loss of binary connections on the network. Since many binary interactions are forbidden, the growth process favours the accumulation of a great number of individual interactions on the allowed links. Increasing the percentage of forbidden links implies a severe damage to network topology, whith a not negligible quantity of interactions being erased, resulting in almost the totality of specialists being connected to one or two supergeneralists. 


\section{Conclusions} \label{sec:conc}
We have presented a novel link-aggregation process for network growth based on individual neutrality. Our results support the hypothesis that an important part of mutualistic networks topology arise as a consequence of random individual encounters and that different species abundance account both for nestedness and truncated degree and strength distributions. Forbidden links do not improve the accuracy of the results but on the contrary they severly damage the network architechture.  This suggests that morphological and phenological restrictions are somehow being accounted for in the probabilistic attachment rule. Its effect on the degree and strength distributions strongly discourages \emph{forbidden links} as candidates for the observed exponential truncations, in favour of other hypothesis such as finite size effect.\\
In conclusion, our individual interaction model succeeds at reproducing the main features of mutualistic networks making use of just three input parameters while remaining simple enough to be analytically solvable. Undoubtedly, claiming to enclose all the  ecological complexity of an ecosystem under such a simple mathematical process would be preposterous. Our model is missing a lot of information about the many processes happening during the development of mutualistic communities. Nonetheless, our results suggest that its underlying mechanism might be capturing some of the key ingredients leading to the formation of mutualistic networks. Further research might lead to more complicated versions of the model aiming to more precise fits to empirical data. These modifications might involve testing the effect of more realistic \emph{birth} functions, considering individual deaths (link deletion) or interespecific competition.

\section{Bibliography}
\bibliography{EMS_arxiv.bib}

\section{Web References}
\begin{itemize}
	\item Interaction Web Database. \url{http://www.nceas.ucsb.edu/interactionweb/}. Last accesed: 2013-06-30. \label{IWDB}
\end{itemize}

\end{document}